\documentclass[aps,pre,floats, twocolumn,noshowpacs,superscriptaddress, longbibliography]{revtex4-2}

\usepackage{graphicx,epsfig}
\usepackage{times}
\usepackage{bm}
\usepackage{lipsum} 
\usepackage{amssymb,amsmath,multirow,rotate,color,mathtools}
\usepackage[utf8]{inputenc} 
\usepackage[dvipsnames]{xcolor}
\usepackage{float}
\usepackage{cancel}

\begin{document}

\title{Comparing temporal and aggregated network descriptions\\ of fluid transport in the Mediterranean Sea } 

\author{Kishor Acharya} 
\thanks{These authors contributed equally}

\affiliation{Instituto de F\'isica Interdisciplinar y Sistemas Complejos IFISC (CSIC-UIB), Campus UIB, 07122 Palma de Mallorca, Spain}
\affiliation{University of Luxembourg, Luxembourg}

\author{Javier Aguilar}
\thanks{These authors contributed equally}

\affiliation{Investigador ForInDoc del Govern de les Illes Balears en el departamento de Electromagnetismo y Física de la Materia e Instituto Carlos I de Física Teórica y Computacional, Universidad de Granada, Granada E-18071, Spain}
\affiliation{Instituto de F\'isica Interdisciplinar y Sistemas Complejos IFISC (CSIC-UIB), Campus UIB, 07122 Palma de Mallorca, Spain}
\affiliation{Laboratory of Interdisciplinary Physics, Department of Physics and Astronomy “G. Galilei”, University of Padova, Padova, Italy}

\author{Lorenzo Dall'Amico}
\thanks{These authors contributed equally}

\affiliation{ISI Foundation, Turin, Italy}

\author{Kyriacos Nicolaou}
\thanks{These authors contributed equally}
\affiliation{Centre for Complex Systems Studies, Utrecht University, Utrecht, The Netherlands}
\affiliation{Cell Biology, Neurobiology and Biophysics, Department of Biology, Faculty of Science, Utrecht University, Utrecht, The Netherlands}



\author{Sandro Meloni}
\affiliation{Istituto per le Applicazioni del Calcolo ``Mauro Picone'', Consiglio Nazionale delle Ricerche, Roma, Italy}
\affiliation{Centro Studi e Ricerche "Enrico Fermi" (CREF), Roma, Italy}
\affiliation{Instituto de F\'isica Interdisciplinar y Sistemas Complejos IFISC (CSIC-UIB), Campus UIB, 07122 Palma de Mallorca, Spain}

\author{Enrico Ser-Giacomi}
\email{enrico.sergiacomi@gmail.com}

\affiliation{Instituto de F\'isica Interdisciplinar y Sistemas Complejos IFISC (CSIC-UIB), Campus UIB, 07122 Palma de Mallorca, Spain}

\date{\today}

\begin{abstract}
 
Ocean currents exhibit strong time dependence at all scales that influences physical and biochemical dynamics. Network approaches to fluid transport permit to address explicitly how connectivity across the seascape is affected by the spatiotemporal variability of currents. However, such temporal aspect is mostly neglected, relying on a static representation of the flow. We here investigate the role of current variability on networks describing physical transport across the Mediterranean basin. We first focus on degree distributions and community structure comparing ensembles of temporal networks that explicitly resolve time dependence and their aggregated, i.e. time-averaged, counterparts. Furthermore, we explore the implications of the two approaches in a simple reaction dispersal model for a generic tracer. Our analysis evidences that aggregation induces structural network changes that cannot be easily avoided, not even introducing a pruning of the aggregated adjacency matrix. We also highlight that, depending on the time scales considered, the importance of the temporal features of the networks can vary significantly. Finally, we find that the tracer evolution obtained from a temporal dispersal kernel cannot be always approximated by aggregated adjacency matrices, in particular during transients of the dynamics.

\end{abstract}

\maketitle

\section{Introduction}

Almost every aspect of the oceanic environment, from physical dynamics to biochemical processes, presents a marked time dependence \cite{stammer1997global, williams2011ocean}. Such temporal variability is both observed at short scales of a few hours \cite{thomas2008submesoscale} as well as at the slower inter-annual climatic ones \cite{levin2015deep}. This is particularly relevant for oceanic currents, that may drive chaotic fluid transport dynamics across the seascape \cite{ottino1989kinematics, vulpiani2009chaos,lacasce2008statistics}, deeply influencing other processes that depend on it, such as tracers redistribution, organisms dispersal or pollutants spreading.

Over the last decade, tools borrowed from network theory brought new insights into the study of transport processes in fluid flows \cite{SerGiacomi2015}. Specifically, in the Lagrangian Flow Network framework, sub-regions of the ocean are represented as nodes of a network, while weighted links between the nodes model the fluid exchange between the regions \cite{SerGiacomi2015,ser2021explicit}. This approach proved its utility for the characterization of physical transport patterns in the ocean and atmosphere \cite{ser2015most,ser2015dominant,ser2021lagrangian} and of marine organisms dispersal \cite{dubois2016linking}. 

Even though Lagrangian Flow Networks can be represented with temporal graphs \cite{ser2015most, thompson2018variability, ser2021explicit}, the common practice is to aggregate i.e. average the temporal component of the network, thereby obtaining a unique static representation of the flow. However, this approximation may be insufficient to describe the dynamical processes unfolding on the evolving network structure at comparable time scales \cite{barrat2008dynamical, iribarren2009impact, delvenne2015diffusion,dallamico2024embeddingbased}. Thus, in the oceanic context, using an aggregated network requires careful consideration of the potential loss of detailed information compared to the time-resolved approach. 

For the past 20 years, researchers in network science developed theories and models to study temporal networks \cite{holme_modern_2015,masuda2016,holme2012}, effectively extending tools and measures proposed for the analysis of static graphs to cope with the time-varying nature of interactions and their characterization. With this work, we use these tools to quantify the differences between static and time-resolved representations of oceanic Lagrangian Flow Networks, assessing their effect on transport and biological processes.     

To this aim, we consider two ensembles of temporal Lagrangian Flow Networks representing surface transport across the Mediterranean Sea and compare them with their equivalent aggregated graphs. 

We limit our analysis to the surface layer of the ocean due its relevance for several processes like air-sea exchanges, plastic pollution, larval transport, plankton dynamics and human activity. To uncover the effect of seasonal variations, the first temporal network is composed of $12$ consecutive one-month snapshots across the same year $2002$, while the second by $10$ networks representing the month of July across $10$ years, from $2002$ to $2011$ to study yearly changes. To compare the temporal and aggregated networks, we first investigated some basic topological measures, such as the resulting degree distributions of the two approaches. We then considered a pruned version of the aggregated network that preserves the average degree of the temporal network. We show that, even if the two networks have the same number of link, this simple pruning it is not able to reproduce the basic features of the temporal snapshots.  Furthermore, we examined the temporal mesoscale organization of the network, observing recurrent patterns in the community structure of the snapshots. Finally, we probed the effects of these structural differences on a minimal reaction-dispersion model simulating the dynamics of a generic tracer.

\section{Methods}\label{sec:methods}

\subsection{Network construction}

Following the approach described in \cite{SerGiacomi2015}, we build different Lagrangian Flow Networks representing fluid transport across sub-regions of the Mediterranean surface for different time windows.

We discretize the Mediterranean basin into a grid of $8196$ equal-area cells of 0.25° latitudinal extension. Cells correspond to the nodes of the network whose edges represent surface transport across the basin \cite{ser2021explicit, SerGiacomi2015,ser2015most}. To quantify horizontal transport, we fill each node with $100$ ideal fluid particles and, integrating realistic velocity fields of surface currents, we reconstruct the Lagrangian trajectories of each particle in time. Velocity fields taken from the Mediterranean Forecasting System (MFS) based on NEMO-OPA (Nucleus for European Modelling of the Ocean-PArellelis, version 3.2 \cite{Coppini2022}). This data-assimilative operational model has been implemented in the Mediterranean at 1/16° degree horizontal regular resolution and we use the first of 72 unevenly spaced vertical levels \cite{simoncelli2014mediterranean, Oddo2009}. Lagrangian trajectories are simulated by integrating horizontal velocities bilinearly interpolated using a Runge-Kutta fourth order algorithm with a time step of $6~$hours. We note that, on the one hand, the network discretization implies a coarse-graining of dynamics that hides patterns smaller than the node size. On the other hand, it introduces spatial diffusion that is needed to parametrize sub-grid scale processes not accounted by the modeled velocity fields.

The set of Lagrangian trajectories is encoded in a transport matrix, denoted $\mathbf{P}\left(t_0, \tau\right)$. Its entry $ij$ is  proportional to the amount of particle exchanged between $i$ and $j$ from the initial time $t_0$ to the final one $t_0+\tau$. The transport matrix can be thus interpreted as the adjacency matrix of a weighted and directed network \cite{SerGiacomi2015,ser2021explicit, ser2020impact}.  We also define the out-degree of node $i$  as the number of edges starting from $i$. Throughout the rest of the manuscript, we refer to the degree distribution $P(k)$ as the probability that a randomly selected node has out-degree $k$.


\subsubsection*{Aggregated network}

Given the set of the matrices $\mathbf{P}\left(t_0, \tau\right)_{i j}$, representing the temporal snapshots, its aggregated counterpart is obtained by averaging across different time windows of duration $\tau$ \cite{SerGiacomi2015,ser2021explicit}. In particular, given an ensemble of $T$ transport matrices $\mathbf{P}\left(t_i,\tau\right)$ with $t_1<\dots<t_T$, the aggregated average matrix is defined as:
\begin{equation}
\label{eq:P_agg}
\bm{\hat{P}}^{\tau}_{i,j}:=\frac{1}{T}\sum_{x=1}^T \bm{P}\left(t_x,\tau\right)_{i,j}.
\end{equation}

\subsubsection*{Pruned network}

\begin{figure}[!t]
	\centering
	\includegraphics[width = 0.8\columnwidth]{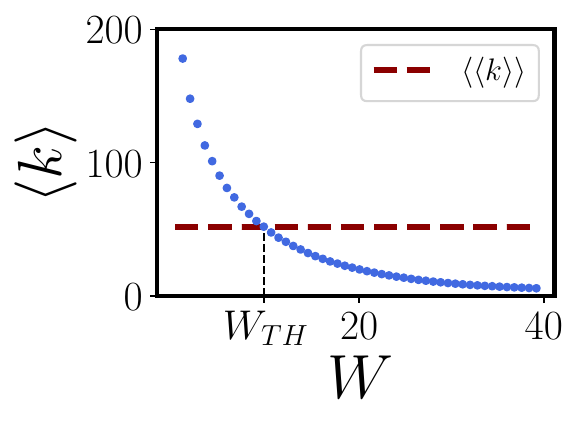}
	\caption{\footnotesize \textbf{Average out degree in the pruned network as a function of the threshold $W$}. The blue dots are the out degree of the aggregated network obtained removing all values smaller than $W$. The horizontal dashed line marks the average network degree averaged over all monthly networks ($\langle \langle k\rangle\rangle\approx 50$). The intersection between the dotted and dashed lines points to the threshold weight ($W_{TH}$) used to build the pruned aggregated network.}
	\label{fig:pruned}
\end{figure}

The aggregation process Eq.~\eqref{eq:P_agg} necessarily generates denser networks, in the sense that the average degree of the average network is always equal or larger than the average degree of single temporal snapshots. This is because the degree only depends on Boolean variables (existence or not of a given edge) and the sum performed in the average converts into an ``or'' operation for the degree of the aggregated matrix. We thus propose a simple modification of the aggregated matrix that both adds and prunes edges to generate a time-averaged network with statistics that better represent single snapshots of the temporal network. We consider a pruning mechanism that consists of removing all the edges from the aggregated matrix that have a weight smaller than a threshold $W$. In Fig.~\ref{fig:pruned} we show how the average degree of the aggregated network decreases for different values of $W$. We choose the threshold pruning weight, $W_{TH}$, to be such that the resulting pruned network's average degree corresponds to the mean of all snapshot's average degree, satisfying
\begin{equation*}
\langle\langle k\rangle\rangle =\frac{1}{T}\sum_{t=1}^T \langle k_t\rangle,
\end{equation*}
where $T$ is the number of snapshots and  $\langle k_t\rangle$ is the average out-degree of the $t^{\text{th}}$ snapshot.

\subsection{Dynamical community detection}

The complexity of spatio-temporal transport patterns in the ocean is reflected in a marked community structure in Lagrangian Flow Networks \cite{SerGiacomi2015,ser2020impact}. To quantify how temporal variations affect the mesoscale organization of the network, we will analyze community structures across the different snapshots and compare them.

Community detection \cite{rossetti2018community, fortunato2016community} is a commonly studied inference problem that consists of partitioning the nodes of a network into tightly and non-overlapping groups. For each node $i$, one wants to define a mapping $i \to \ell(i) \in \{1, \dots, q\}$, where $q$ is the number of communities. A practical challenge in this setting is related to the complexity of the network structure that is weighted, directed, and temporal. We use a dynamical spectral clustering algorithm inspired by \cite{dall2020community}, adapted to weighted and directed graphs.

\medskip

The core idea of spectral clustering is to represent each node of the network as a vector in a low dimensional space, using the eigenvectors of a suited graph matrix representation. The vectors can then be divided into groups with, for instance, \emph{k-means} algorithm \cite{lloyd1982least} of \emph{expectation-maximization} \cite{dempster1977maximum}, as we do in the following. For static, weighted, and directed networks, ref.~\cite{coste2021simpler} showed that, even in the sparse regime in which typically many spectral algorithms tend to fail, one can obtain this embedding computing the $k$ largest eigenvalues of the weighted adjacency matrix, store them in the columns of an embedding matrix $X \in {R}^{N\times k}$ and interpret its rows as embedding vectors. For the temporal setting, we generalize this technique by building the following matrix $\mathbf{M} \in R^{NT \times NT}$: 

\begin{align*}
\mathbf{M} = \begin{pmatrix}
\mathbf{P}(t_0, \tau) & h I_N & \mathbf{0} & \dots & \mathbf{0}\\
hI_N & \mathbf{P}(t_1,\tau) & hI_N & \dots & \mathbf{0}\\
\mathbf{0} & hI_N & \mathbf{P}(t_2, \tau) & \dots & \mathbf{0} \\
\vdots & \vdots & \vdots & \ddots & \vdots\\
\mathbf{0} & \mathbf{0} & \mathbf{0} & \dots & \mathbf{P}(t_T,\tau)
\end{pmatrix},
\end{align*}
where $I_N$ is the diagonal matrix of size $N$ (the number of nodes) and $h > 0$ is a regularization parameter that imposes that the community label of each node must change slowly across time. In our simulations, we set this regularizer value to $h = 0.2~p$, where $p$ is the sum of all entries of $\mathbf{P}$. Every row of this matrix, and consequently of its eigenvectors, is associated with a node at a given time instant.
The derivation of $\mathbf{M}$ can be obtained following the steps described in \cite{dall2020community}, computing the Hessian matrix of the na\"ive mean-field free energy (instead of the Bethe free energy) related to the same Hamiltonian matrix. 

On top of the flexibility that allows one to deal at once with weighted, directed, temporal, and sparse graphs, a major advantage of this approach is that one can estimate the number of communities $k$ in an unsupervised fashion. In fact, $\mathbf{M}$ is non-Hermitian, hence its eigenvalues are generally complex. In a model-based approach, however, the largest of these eigenvalues, are associated with its expectation that is -- by definition -- supposed to be symmetric of rank $k$. Hence, the first $k$ largest eigenvalues of $\textbf{M}$ are real and by identifying the position of the first complex eigenvalue one can also estimate the number of communities, a typically hard task to solve.

\subsection{Minimal reaction dispersal model}\label{sec:met_reactive_model}

Here we introduce a minimal model for reaction dispersal dynamics happening on top of the network. We aim at modeling the evolution of a generic tracer concentration field \cite{Emiliobook} across the Mediterranean that is affected by the intrinsic reactive process as well as dispersal due to ocean currents. We start defining a concentration tensor vector for which each entry $n_i(t)$ corresponds to the concentration in the node $i$ of the network at the discrete time $t$. Then, we assume that the tracer concentration at a given node is redistributed by currents and experiences growth and decay. We model such processes advecting the tracer concentration by left-multiplying $n(t)$ with the network adjacency matrix $\mathbf{P}$ and implementing simple logistic dynamics. We can thus write the concentration in node $i$ at time $t+\tau$ as: 
\begin{align}
\label{eq:model_dynamics}
n_i(t+\tau) = (\mu + 1) &\sum_j \mathbf{P}_{ji}(t,\tau)n_j(t) \nonumber \\
 &-\frac{\mu}{K} \Big[ \sum_j \mathbf{P}_{ji}(t,\tau) n_j(t) \Big]^2,
\end{align}
where $\mathbf{P}_{ji}(t,\tau)$ are entries of the row-normalized adjacency matrix of the transport networks to be considered. The parameters $\mu$ and $K$ are the growth rate and the carrying capacity of the logistic reactive dynamics.

We also introduce three metrics to characterize the spatial patterns of the tracer field $n_i(t)$ across time. We first define the \emph{fraction of colonized nodes} as the proportion of nodes that present a non-zero concentration at a given time $t$. Then, being $M$ the number of nodes in the network, we denote the \emph{spatial mean} of $n_i(t)$ as
\begin{equation}
\frac{1}{M} \sum_{i=1}^M n_i(t)
\end{equation}
and its \emph{spatial standard deviation} as
\begin{equation}
\sqrt { \frac{1}{M}  \sum_{i=1}^M n_i^2(t) - \bigg ( \frac{1}{M}  \sum_{i=1}^M n_i(t) \bigg)^2} .
\end{equation}

\section{Results and Discussion}

We here discuss the main results of our analysis, which is based on comparing the degree distribution, the community structure, and the population dynamics on the temporal and aggregated transport networks. 

\subsection{Properties of degree distribution}

\begin{figure*}[!t]
	\centering
	\includegraphics[scale=0.45]{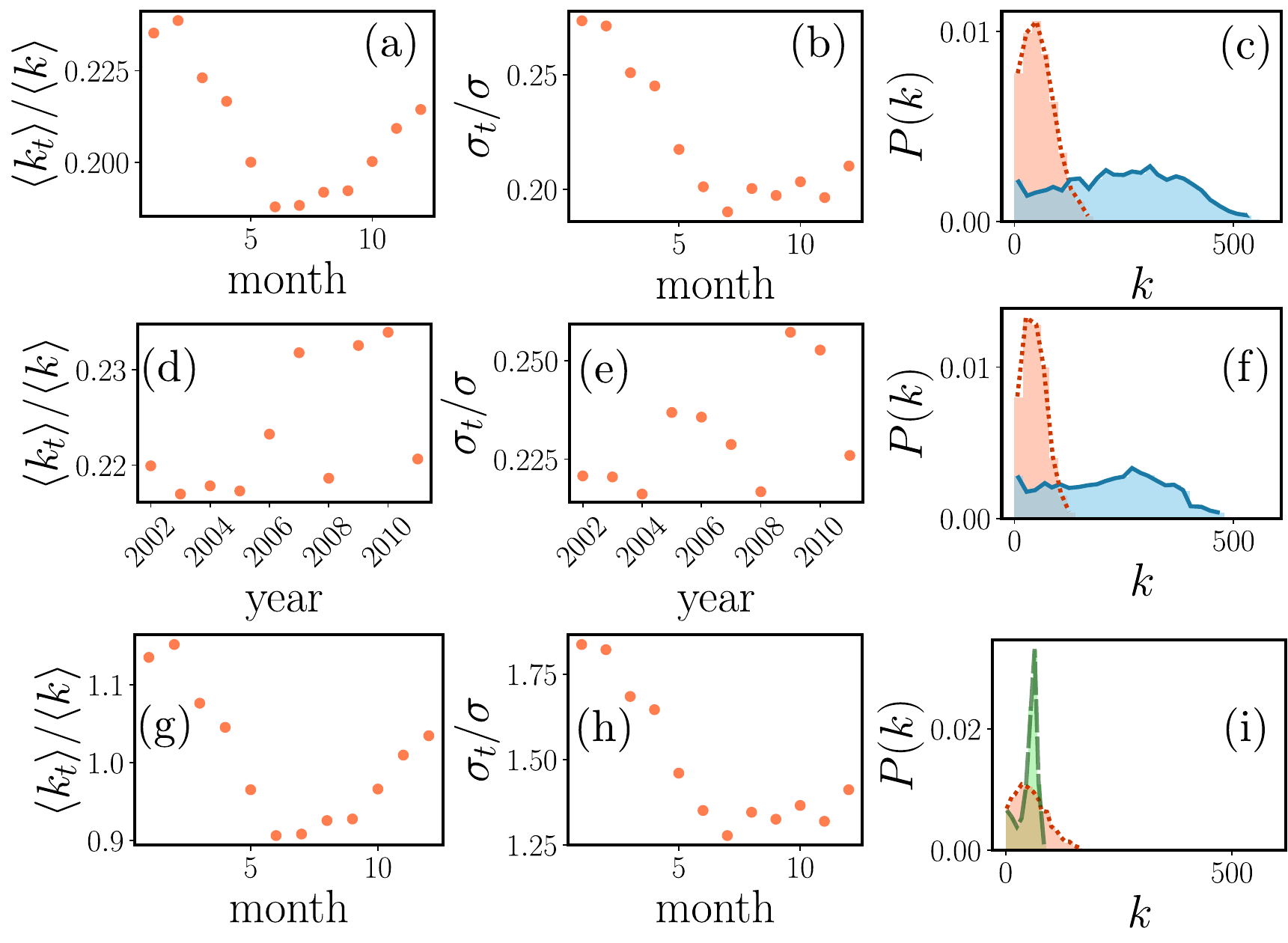}
	\caption{\footnotesize\textbf{Degree distribution of the aggregated and temporal transport matrices} \emph{Panel (a)}: average degree of the transport networks at different months, $\langle k_t\rangle$, divided by the average degree of the aggregated matrix ($\langle k \rangle\approx 250$). \emph{Panel (b)}: similarly to (a), standard deviations of the average degree of the temporal networks $\sigma_i$ and of the aggregated one ($\sigma\approx130$). \emph{Panel (c)} : degree distribution corresponding to one snapshot (dotted red line) and the degree distribution for the aggregated network (continuous blue line). \emph{Panels (d), (e), (f)}: similar to panels (a), (b) and (c) but for the interannual temporal networks. \emph{Panel (g)}: similar to (a), but average degree of snapshots is divided by the average degree of the pruned network ($\langle k \rangle \sim 50$). \emph{Panel (h)}: similar to (b), but standard deviation of snapshots is divided by the standard deviation of the pruned network ($\langle \sigma \rangle \sim 20$).  \emph{Panel (i)}: degree distributions corresponding to the pruned network (dashed green line), and one monthly network corresponding to January (dotted red line). }
	\label{fig:month_by_month_vs_aggregated}
\end{figure*}
\begin{figure*}[!t]
	\centering
	\includegraphics[width = 0.65\linewidth]{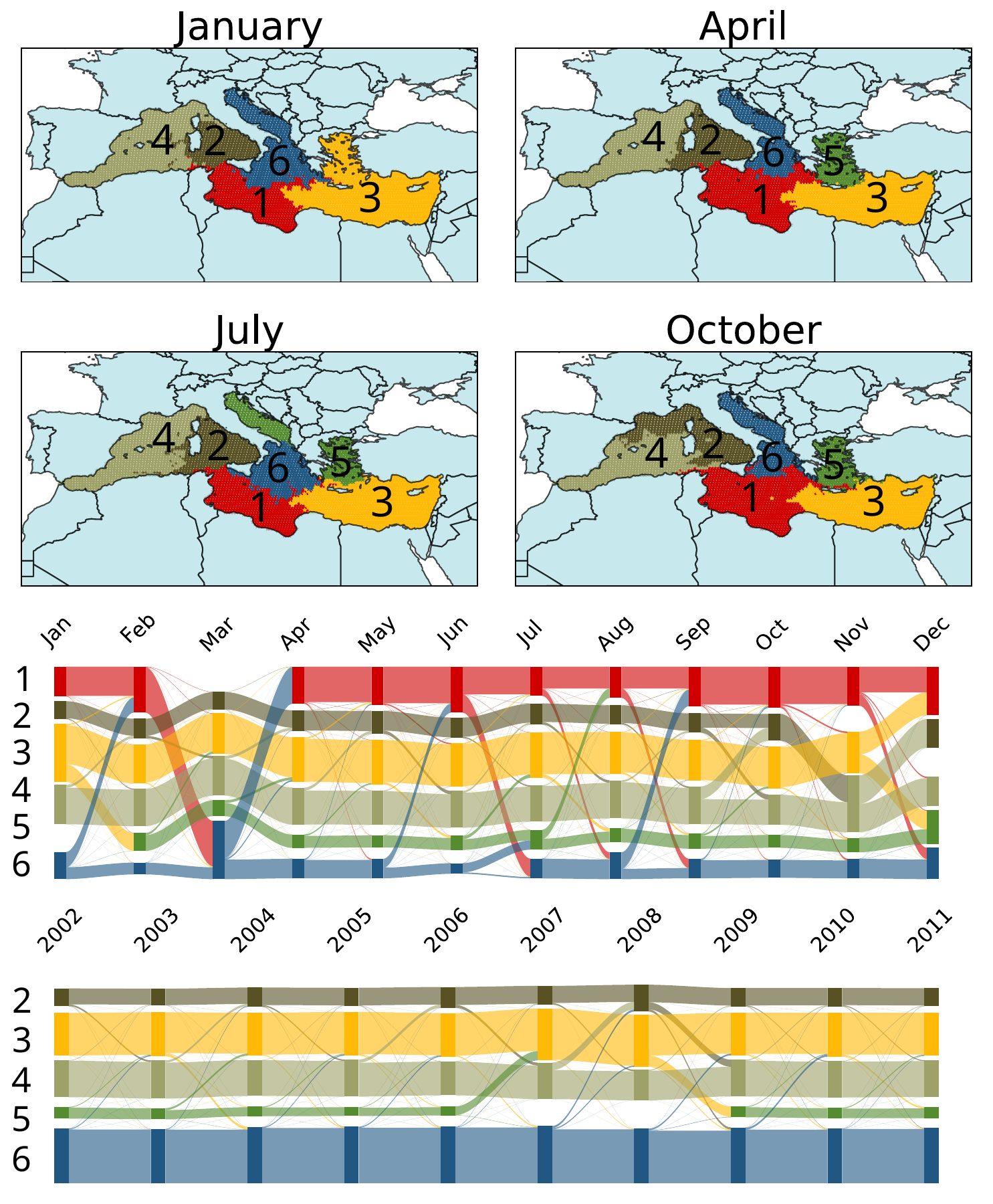}
	\caption{\footnotesize\textbf{Dynamical community detection on the Temporal Flow Networks}. Community detection is performed on the transport networks and $6$ communities (represented by the colors) are detected. The four panels in \emph{(a)} show the results for four monthly transport networks (Jan, Apr, Jul and Oct 2002). \emph{Panel (b)}: Sankey plot of the community structure evolution over time for the by inter-annual graph. Each column refers to a snapshot (here a month of a different year) and the height of the bars are proportional to the number of nodes in a given community. The flows connecting the bars are associated to nodes changing their label across time. \emph{Panel (c)}: same as in Panel (b) but referred to the seasonal data. Here each snapshot corresponds to a month in the same year. The plots in Panel (a) refer to the same community structure reported here.}
	\label{fig:map}
\end{figure*}

\subsubsection*{Seasonal data  }\label{sec:Global_properties_month_by_month} 

We first investigate the differences in general statistical properties of the temporal and aggregated networks. In Fig.~\ref{fig:month_by_month_vs_aggregated}, we compare the global features of 12 month-by-month matrices and their corresponding aggregated matrix.
Fig.~\ref{fig:month_by_month_vs_aggregated}-(a) shows that, as expected, the average degree of the aggregated matrix is always larger than the average degree of any of the twelve month-by-month matrices, making the aggregated matrix significantly denser. We note that the difference between the density of single snapshots and of the aggregated matrix is due to changes in connectivity within snapshots i.e. links that appear and disappear across the temporal sequence. We thus conclude that the snapshots links geometry undergoes several changes during the observation time window. Fig.~\ref{fig:month_by_month_vs_aggregated}-(b) shows that also the degrees standard deviation is larger in the case of the aggregated matrix, meaning that there is an overall increase of the second moment of the aggregated degree distribution. Furthermore, both in (a) and (b) we observe n temporal pattern that we associate with seasonality which is not accounted for in the aggregated matrix.
To further investigate connectivity fluctuations, we plot in Fig.~\ref{fig:month_by_month_vs_aggregated}-(c) the degree distribution of the aggregated matrix together with one of the temporal snapshots. We observe a neat discrepancy between the two distributions, meaning that the aggregation process does not mean a trivial homogeneous shift in the degrees of the temporal matrices. Indeed, the degree distribution of the temporal snapshots peaks around the mean of the distribution, while the aggregated distribution is much flatter. Therefore, in the aggregated network small degrees can occur with a similar probability as much larger degrees. Based on this observation, we assume that while the destinations of some links change significantly between snapshots, the links of other nodes may remain relatively constant through the one-year evolution.

\subsubsection{Interannual data}

We replicate the analysis conducted in section~\ref{sec:Global_properties_month_by_month} with the dataset describing ocean transport in the same month but across ten different years (2002-2011). In the plots in Fig.~\ref{fig:month_by_month_vs_aggregated}-(d-f), we observe a similar behavior to the monthly dataset with differences observed in the degree distribution of the different temporal snapshots and the aggregated matrix, corresponding to bigger connectivity and degree variability in the latter.

\subsubsection{Pruned network}

The pruned network reduces both the variability of degrees and the average degree to values that are closer to the single temporal snapshots, as we expected [see Fig.~\ref{fig:month_by_month_vs_aggregated} (g) and (h)].  However, the degree distribution for the pruned network is bimodal and presents a sharp peak, making it very different from the typical degree distribution of the temporal network [see Fig.~\ref{fig:month_by_month_vs_aggregated} (i)]. Therefore, the snapshot aggregation generates deep structural changes that cannot be removed from a simple pruning of the less connections.

\begin{figure}[!t]
	\centering    
	\includegraphics[width=\linewidth]{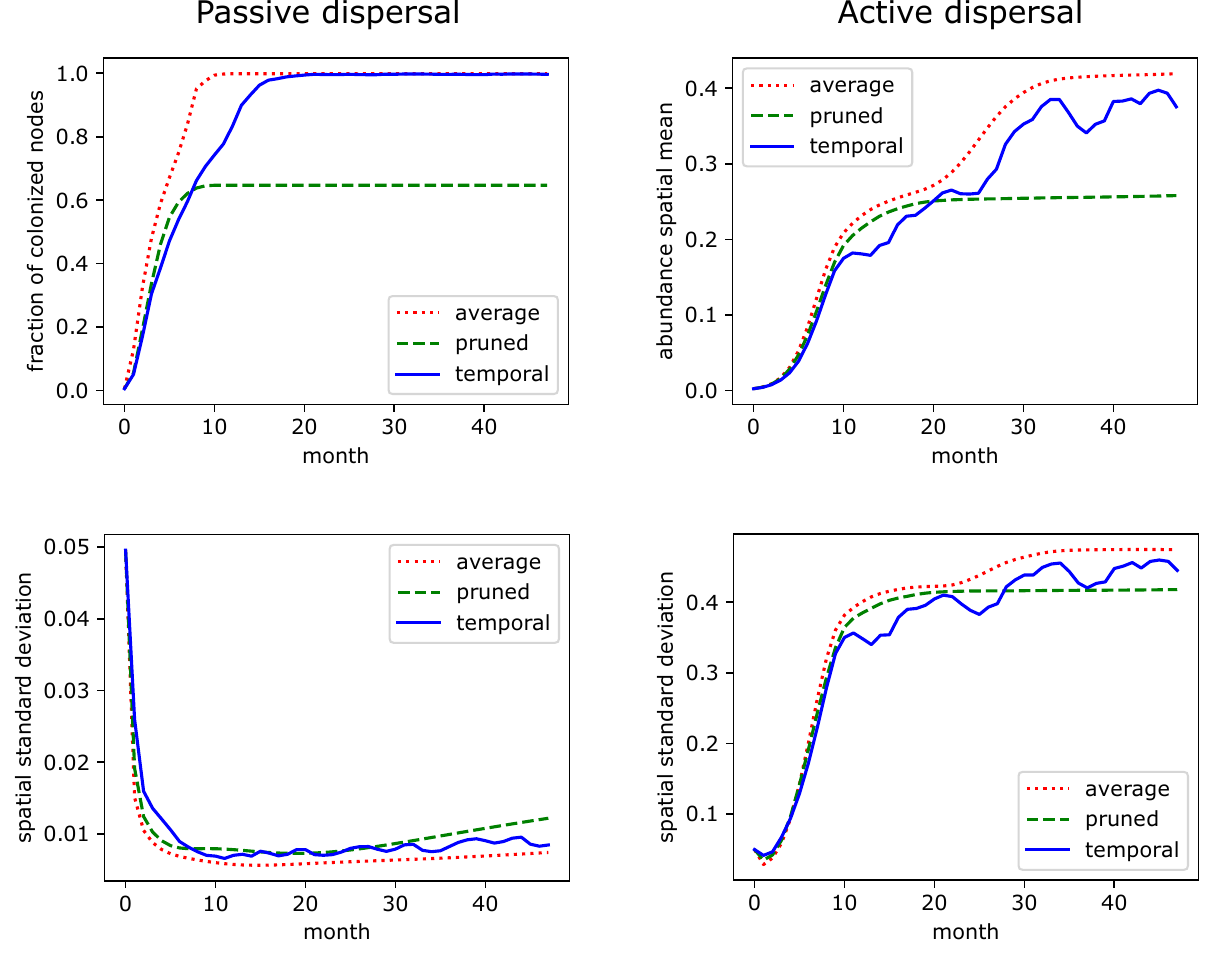}
	\caption{\footnotesize\textbf{Reaction dispersal model outputs}. We show the temporal evolution of the fraction of colonized nodes, spatial mean and spatial standard deviation for both the passive (left column) and reactive (right panel) model setups. In each panel we plot the results obtained using as adjacency matrix the full temporal (blue), the simple average (red) and pruned-average (green). }
	\label{fig:reactivemodeloutput}
\end{figure}

\subsection{Community structure}

Community structures in networks denote groups of nodes that are more densely connected among them with respect to the rest of the network. For Lagrangian flow networks this translates into areas of the sea well mixed internally but with little exchange among them, with potentially crucial consequences for the biological processes taking place inside. 
To understand how these communities are influenced by seasonal (short term) and interannual (long term) variability, we apply our temporal community detection algorithm to quantify temporal changes between snapshots. In the top panels of Figs. \ref{fig:map}a, we plot the community structure in color code for $4$ months and show that, while there is some level of persistence,  community structures are evolving over different months. We remark that some communities have a rather small size (most of them reflecting shallow oceanic regions such as continental shelves), and their variability is shaped by inter-seasonal flows as evidenced by the Sankey plots in Fig.~\ref{fig:map}c. On the other hand, communities across different years are very stable, as shown in Fig.~\ref{fig:map}b, highlighting once more the relevance of seasonal flows for the network structure.

To conclude, this implies that a dynamical process running on the network at the time-scale of months is expected to have significant differences if one uses the temporal or the aggregated network. On the opposite, a less prominent role of time should be observed at the scale of years.

\subsection{Reaction dispersal dynamics}

To conclude our analysis, we simulate the time evolution of a reaction dispersal process on the network using the model introduced in Section \ref{sec:met_reactive_model}. Our approach does not have the ambition of providing a realistic description of tracer dynamics, but it aims to understand if the temporal nature of the dispersal kernel can generate different spatiotemporal patterns of a generic concentration field, even assuming an extremely simplified dynamics. To this end, we model the evolution of the concentration tensor $n_i(t)$ across 48 consecutive months. We initialize its entries to 1 for a subset of 20 contiguous nodes centered in the middle of the Central Mediterranean Sea, setting the remaining ones to zero. We run the dynamics defined in Eq. (\ref{eq:model_dynamics}) with three independent settings,  each using a different adjacency matrix: the temporal one, the simple averaged, and the pruned network. Moreover, we also compare the case of a purely passive dispersal process, setting $\mu=0$, with the one of a reactive dispersal with $\mu=K=1$. 

In Fig. \ref{fig:reactivemodeloutput}, we show different spatial statistics of $n_i(t)$ across time for the above-mentioned model settings. We observe that, when the temporal adjacency matrix is used, the time variability of the statistics computed increases. For the passive dispersal case, the fraction of colonized nodes (i.e. nodes with non-zero concentration values) reaches 1 with different speeds in the temporal and the average settings, while it does not saturate to 1 when using the pruned matrix. This is because the network pruning creates disconnected components in the network making impossible the connection between every pair of nodes. We also note that the spatial standard deviation initially declines but after 10-20 months starts to slowly increase. This phenomenon is explained by the presence of sets of nodes acting as sinks that accumulates tracer concentration over time. We see qualitative and quantitative differences also in the reactive setups of the model, both for the spatial mean and standard deviation of the concentration fields. 

Overall, in both passive and reactive cases, we observe that the tracer dynamics obtained from an temporal dispersal kernel cannot be always approximated by either the simple averaged or the pruned-average adjacency matrices. We also note that the difference between the temporal and the aggregated approaches seems to be more relevant during transients of the dynamics affecting the speed at which quasi-asymptotic states are reached. However, exactly this transient dynamics could play a crucial role for biological processes as it unfolds at a comparable time-scale.

\section{Conclusions}

In this work, we analyzed temporal networks describing time-dependent transport processes in the ocean and compared them with their time-averaged counterparts. Analyzing their topology, we observe that both the total number of links and the heterogeneity of nodes degree is bigger in the aggregated temporal networks than for single temporal snapshots. While this is expected, as the number of links can only increase through the aggregation process, the variability increase is our first indicator of deep structural differences between the aggregated and temporal networks. The pruned network, designed to reduce the number of links to values comparable to those observed in single snapshots, still exhibits significant differences from the degree distribution of the temporal network. Therefore,  we can conclude that aggregating temporal snapshots induces some deep structural changes that cannot be easily avoided.

We observed that the temporal scales play a crucial role when evaluating the significance of time dependency in oceanic transport. While temporal changes within communities are apparent on a monthly basis, they become less evident when examining the network's evolution over the years. We explain the difference between the across-month and across-year evolution as a dominance of seasonality over interannual variability. This is consistent with a seasonal scale dominated by the solar cycle and the consequent strong changes of the heat input toward the ocean and weather configuration patterns. The interannual scales are instead influenced by climate dynamics, essentially ENSO and NAO, and their effects on the Mediterranean circulation are more elusive.

Our minimal reaction-dispersal model shows that the structural differences pinpointed so far have an impact on the predicted evolution of tracers. In particular, aggregated networks tend to produce faster advection patterns. This effect has already been observed in epidemiological modeling, in which denser aggregated networks artificially speed up the spreading of an infectious disease~\cite{kobayashi2019structured}.

To summarize, the preferred option to build faithful descriptions of oceanic transport within a year should rely on a temporal network framework. However, due to a lack of data or computational resources, this might be unfeasible depending on the application. In such cases, one should bear in mind that averaging networks is a subtle task, which might require sophisticated methods to prevent structural network changes during aggregation. Thus, future work should be directed towards building and refining methods that can aggregate networks while preserving structural properties (e.g.~\cite{kobayashi2019structured}). We also envisage that this work can motivate theoretical studies aimed at parameterizing the temporal dependence of oceanic transport networks and building effective generative models.

\acknowledgments 

We acknowledge Johnny Tong for his contribution to the research during its early stages. This work is the output of the Complexity72h workshop, held at IFISC in Palma de Mallorca, Spain, 26-30 June 2023 \href{https://www.complexity72h.com}{complexity72h.com}. LD acknowledges the support from Fondation Botnar ((EPFL COVID-19 Real Time Epidemiology I-DAIR Pathfinder) and from the Lagrange project of Fondazione CRT. Partial financial support has been received from the Agencia Estatal de Investigaci\'on and Fondo Europeo de Desarrollo Regional (FEDER, UE) under project APASOS (PID2021-122256NB-C21/PID2021-122256NB-C22), the María de Maeztu project CEX2021-001164-M, funded by the  MCIN/AEI/10.13039/501100011033, and the Conselleria d'Educaci\'o, Universitat i Recerca of the Balearic Islands (Grant FPI FPI\_006\_2020), and the contract ForInDoc (GOIB). SM acknowledges support from the project `CODE - Coupling Opinion Dynamics with Epidemics', funded under PNRR Mission 4 `Education and Research' - Component C2 - Investment 1.1 - Next Generation EU `Fund for National Research Program and Projects of Significant National Interest' PRIN 2022 PNRR, grant code P2022AKRZ9. E.S-G. acknowledges support from grant PID2021-123352OB-C32 funded by MICIU/AEI/10.13039/501100011033 and FEDER, `Una manera de hacer Europa'.


%

\end{document}